\documentstyle[12pt]{article}
\input epsf
\title{Gauge Physics of Finance: simple introduction}
\author{ Kirill Ilinski\thanks{E-mail: kni@th.ph.bham.ac.uk} \\ [0.2cm]
{\small\it School of Physics and Space Research, University of
Birmingham,} \\ {\small\it Edgbaston B15 2TT, Birmingham, United
Kingdom} }

\begin{document} \maketitle \vskip -9.5cm \vskip 9.5cm

\begin{abstract}
In this paper we will state the fundamental principles of the gauge
approach to financial economics and demonstrate the ways of its
application. In particular, modeling of real pricing processes will be
considered for an example of S\&P500 market index. Derivative pricing and
portfolio theory are also briefly discussed.
\end{abstract}

\section{Introduction: Why is it financial physics that has been
tackled?}

Experts say that people rest when they contemplate symmetrical figures.
One of the ways to explain this phenomenon is that a symmetrical picture
is easy to be "grasped" by the eye, which makes an impression that it is
comprehensible and easy to control as it can be decomposed into simpler
ideal figures that are well known and have been recognizable since one's
childhood. This discourse can be applied to the cognition process in
general: cognizing new things we "decompose" them into fundamental
components ("blocks") and establish links between them. The more
fundamental components there are at our disposal, the more beautiful and
comprehensive the real picture is.

During its long period of existence physics has accumulated a wide range
of "blocks" that are the building blocks of an edifice of contemporary
knowledge of the physical world. These "blocks" are special since they
have been specially selected for describing sophisticated systems
consisting of a great number of elements that interact with each other.
That is why it is not no wonder that these "blocks" are made use of in
other areas besides physics that are associated with response to the
impact of many interrelated factors. Urban development, traffic jams,
economic issues are only several examples of such applications.

Like any other branch of science designed to provide quantitative
description, physics uses the language of mathematics. However, there is
a distinctive difference between mathematics and physics. This difference
lies in the way of setting a task. As we advance in tackling the subject
of our studies -- financial modeling -- we will consider how financial
mathematics is different from financial physics -- a branch of science
that is just emerging. To put it in a nut shell, financial mathematics
deals with the question "How?". For instance, how to form an effective
investment portfolio or how to assess and hedge derivatives? Financial
physics deals with the question "Why?". Below the three most important
"Whys" we are going to deal with further are listed:
\begin{enumerate}
\item Why do we
observe certain statistical characteristics of real pricing charts?
\item Why do traders analyze pricing charts and use technical analysis
that is discarded by all economists?
\item Why does the Black-Scholes
equation not always describe real prices?
\end{enumerate}

It may seem that a practitioner is more interested in "How?" rather than
in "Why?". However, from history it is known that after the question
"Why?" has been answered, there is an answer to the question "How?" as
well. For example, in contemporary financial mathematics almost all
answers to the question "How?" are derived from the answer to the
question "Why?": because long-term opportunities to gain no-risk returns
greater than the returns from a bank deposit should not exist, that is,
because arbitrage opportunities should not exist in a market of rational
agents. The latter statement leads to a hypothesis of an efficient
market, a model of stochastic processes for price description,
martingales and other tools of financial mathematics. This is the basis
for the huge industry of quantitative analysis and financial economics.
It is the no-arbitrage assumption that financial physics is going to
attack, by accepting the existence of  short-life arbitrage
opportunities.

Let us look into the no-arbitrage condition closer. A great number of
factors affect prices constantly, ranging from fundamental information
release (which happens comparatively rarely) to the appearance at the
market of an investor ready to make a timely deal which can change
quotations. Price changes cause situations when some assets are more or
less attractive than other assets with the same level of risk. The
arbitrage  (or no-risk returns) is only a particular example of the
described above mispricing. It is clear that such a situation  can not
last for long, and rational investors (or almost rational) will smooth
out the situation by buying profitable and selling unprofitable assets
(let us put aside for now bubbles and crashes), after which the return on
assets will again reflect their level of risk properly. In other words,
the arbitrage opportunities will cease to exist. Thus, the no-arbitrage
assumption may be considered as a statement of the instantaneous
character of the smoothing described above and of the infinite speed of
money flows. But is it really so? Provided that there are market
imperfections and the money flows do not transfer instantly, to what
extend can the no-arbitrage picture be justified? Can money flows be
disregarded?

It is here that we tackle financial physics. Technically, a solution of
any physical problem boils down to realization of what can be disregarded
without distorting the picture and keeping the most essential details. On
the other hand, all physics "turns" around the transfer of matter or
energy, forces engendering these processes, non-equilibrium dynamics and
equilibration process. Continuing this chain, it can be said that all
financial physics consists of the answers to the questions about money
flows disregarding whether they are described with the help of the gauge
theory (which will be tackled later) or the theory of critical phenomena
or the computer modeling of behavior of a great number of
agents-investors. Summarize all said with the first formula in the
article, financial physics can be defined as
\begin{equation}
\mbox{ \bf Financial Physics = Financial Mathematics + Money flows}
\end{equation}
Further, we are going to demonstrate how the tools of
modern theoretical  physics, its images and objects, physical "blocks"
are used to build a theory describing the short-term ("fast") dynamics of
money flows and addressing three questions "Why?" of financial physics.
We will see that there is a certain similarity between gauge theories  of
fundamental interactions and financial economics arising from general
beautiful symmetries and mathematical (differential geometrical)
structures. Such a similarity essentially limits the number of scenarios
of mathematical description and leads to theories that have been studied
intensively by physicists during the recent 50 years. In this approach,
excess return plays the role of a force field (an analogue of an
electromagnetic field), and money flows created by it play the role of
charge currents. Uncertainty immanent to the financial market becomes an
analogue of quantization inducing uncertainty into physical theories,
while transaction costs turn into an effective charge mass resulting in
inertia and limited time of the money flows reaction. These analogies as
may seem futile at first glance come easily from the mathematical point
of view. Studying price statistics in such a model we achieve the picture
which is very close to a real one, and the first question "Why?" is
successfully answered. The second "Why?" will be addressed when we have a
closer look at how the system develops within a very short period of time
when casual fluctuations have not yet "washed out" the trend. In this
case, the movement of prices and money flows is determined by a system of
equations resulting in indicators developed  in technical analysis and
used by traders. Adjustments to the Black-Scholes equation will appear
naturally when we reformulate financial derivative pricing policy in
terms of a new theory. Further, we will consider each of these situations
and discuss them in more details.

It should be understood that financial physics like any other economic
theory can not be taken for a panacea or Holly Grail. However, it is a
step on the way to deeper understanding and more perfect theory. One
should not expect to have the same number of quantitative coincidences as
in Quantum Electrodynamics. It goes without saying that the laws
regulating the microworld and galaxies are different from those
regulating the people's society, and they can not be interchanged
mechanically. On the other hand, it is impractical to let our prejudices
stop us making use of evident analogies and the experience acquired in
physical sciences. Moreover, besides physical methods there is also
physical methodology, in which one assumes that unless an experiment
negates the theory it has the right to exist.

\section{Analogy with Electrodynamics}

We will start with a simple example. Let us assume that the spot exchange
rate of a dollar to a pound is $F(t)$ at time $t$ while their respective
interest rates are $r_1$ and $r_2$. Assuming that all relevant
information that can effect the exchange rates between $t$ and $t+dt$ is
known and is reflected in these rates we ask how $F(t)$ and $F(t+dt)$ are
interrelated.  It is easy to find that this interrelation can be
expressed by a simple equation:
\begin{equation}
F(t) (1+r_2) = F(t+dt) (1+r_1)
\end{equation}
Let us suggest that for some reasons the right side of the equation is
greater than the left one. This will immediately result in a response
from "smart money" that will borrow pounds at time $t$, exchange them
immediately into dollars, deposit the dollars until time $t+dt$ at an
interest rate $r_1$, after which exchange them back to pounds. The
assumption that the right side of equation (2) is greater than the left
one will ensure a no-risk profit from an arbitrage transaction described
above. However, this condition can not last for long: since few sellers
will choose to sell dollars at time $t$ and few buyers will choose to buy
dollars at time $t+dt$ at initial prices, the exchange rates will change
until equality (2) is redeemed. It is also easy to make certain that the
left part of equation (2) can not stay greater than the right one for a
long time. Equation (2) expresses the condition of absence of arbitrage.
However, we are more interested in the restoration process rather than in
equation (2) itself.

The process of restoration or, as we are going to refer to it, relaxation
will take some time the length of which is determined by the market
liquidity as well as market imperfections such as transaction costs and
the bid-ask spread. The same factors will define the speed of the
relaxation. For instance, the relaxation goes faster when the deviation
from a balanced price is great, which attracts a large number of
arbitrageurs despite transaction costs compared with cases when
deviations are little and transaction costs make transactions profitable
only for big arbitrageurs. The closer the market is to perfection, the
less time relaxation takes and the higher its speed is.

Using this example we can identify two issues that will play a very
important role in further studies.

\subsection{Arbitrage and paths}

Arbitrage or gaining returns from mispricing are always associated with
the flow of assets along two different routes having a common beginning
and end. In the previous example we have compared route (a1): pounds at
the time $t$ $\rightarrow$ dollars at the time $t$ $\rightarrow$ dollars
at the time $t+dt$ $\rightarrow$ pounds at the time $t+dt$ with route
(a2): pounds at the time $t$ $\rightarrow$ pounds at the time $t+dt$.
Returns gained from these two routes of transactions are expressed by
equation (2) assuming there is no arbitrage. Similarly, a pair of routes
(b1) and (b2) can be introduced which begin and end in dollars.

Instead of two routes, a closed path of the assets flow can be studied
which follows the first route from the starting point to the end, and the
second route – from the end to the starting point. In our case it will be
a cyclic path (c): pounds at the time $t$ $\rightarrow$ dollars at the
time $t$ $\rightarrow$  dollars at the time $t+dt$ $\rightarrow$ pounds
at the time
$t+dt$ $\rightarrow$ pounds at the time $t$. Having assigned to each
segment of this path a respective exchange or interest factor (assuming
that each segment that flows backwards has an inverse factor) and having
multiplied these factors along the path and subtracted one we will get an
equation:
\begin{equation}
R(c) = F^{-1}(t) (1 +r_2)^{-1} F(t+dt) (1+r_1) - 1
\end{equation}
which is equal to the discounted profit from an arbitrage transaction
when "cash" follows route (a1), and debts follow route (a2). Further, we
will use a term "excess return on the arbitrage operation" to define this
value.

Besides a cyclic path (c) there is another cyclic path (-c) which is
derived from (c) by changing the flow direction. This path can be also
described by an equation:
\begin{equation}
R(-c) = F(t) (1+r_2) F^{-1}(t+dt) (1+r_1)^{-1} -1
\end{equation}
which is equal to the discounted return on an arbitrage operation when
"cash" follows route (b1), and debts follow route (b2). Combining
equations (3) and (4) one can obtain the following value:
\begin{equation}
R=R(c)+R(-c)
\end{equation}
defining an opportunity to
carry out a (certain) profitable arbitrage operation. Quantity $R$ is not
negative and is equal to zero only if there is no arbitrage. In this
case, equation (2) is equivalent to the equation $R=0$. It is more
convenient to use value $R $rather than equations (4) and (5) separately
especially when we do not know which particular operation is profitable.

\subsection{   Charges and Forces}

Let us go back to a mechanism of establishing balance described in a
paragraph after formula (2). Speaking in general terms we can conclude
that "cash" flows from undervalued assets in overpriced assets and
"debts" flow backwards, so if "cash" flows like charged particle
experiencing a force then "debts" behave like particles with opposite
charge. What is more, flowing in such a way assets make this very force
change diminishing its value. In physics this effect is referred to as
screening. Thus, we may conclude that a financial system behaves in the
same way as a system of charges in a force field which is created and
changed by these charges. Applying physical "blocks" we can determine
that a financial system "cash"-"debts"-arbitrage seems to look like
classical electrodynamics though not in conventional three-dimensional
space but in strange discrete space (time remained unchanged). In our
example with dollars and pounds this new financial space consists of two
points only -- a point "dollars" and a point "pounds" -- where assets
"jump" from one point into another.

At first glance such an analogy with electrodynamics may seem purely
superficial for economists as well as physicists. Frankly speaking, there
are several ways of writing an  equation describing migration and
screening, electrodynamics being just one alternative. That is why
without additional arguments favoring electrodynamics seems artificial
and unjustified. Such additional arguments are provided in the form of
general powerful symmetry which singles out electrodynamics from a great
number of other competitive theories.

Let us study equations (3) and (4) in more detail. It is easy to find
that they have certain remarkable properties: they do not change when
currency units and their respective exchange and interest rates are
changed. For instance, we assume that between time $t$ and $t+dt$ it was
decided to use pence instead of pounds. This decision will not affect the
dollar's interest rate and exchange rate at time $t$, but it will
diminish the exchange rate 100 times at time $t+dt$ (when a pound will
cost 100 new units -- pence) and increase the interest factor 100 times
-- having deposited one pound one will get $100(1+r_2)$ pence. Factor 100
will vanish and equations (3) and (4) will remain unchanged. We could
have applied such consideration to simultaneous change of both dollars
and pounds, which would not alter the result -- equations (3) and (4)
based on closed paths do not change. Going further and applying this
approach to any traded (exchangeable) assets one can see that equations
like (3), (4) and (5) remain unchanged when the assets units scale
changes disregarding whether it is currency, bonds or stocks. In the
first case the change in the scale would mean denomination, in other
cases – the change of the traded lot or merge and split.

Let us suggest now that we are trying to construct a theory that has the
property of not changing when the units of measurements are arbitrary
chosen, that is, it does not depend on the choice of the financial assets
units – the currency nominal value, the lot size, etc. It is no doubt
that the real world has this property, at least to certain extent –
agents do not start behaving in a different way only because they are
dealing with 100 pence instead of pounds or if there are only 50 shares
in the lot instead of 100. In building this theory one can only use such
mathematical objects that remain unchanged when the units of measurement
are changed, for instance, equations (3), (4) and (5). We can prove that
in this case the simplest nontrivial theory will be an equivalent of
electrodynamics! It is this very property that distinguishes
electrodynamics from a great number of other competitive theories.

Building a model theory we did not start from searching for symmetry of
this theory just by chance. We were aided by the methodology of physics.
All contemporary physical theories are built based on the symmetry
inherent in them. The general theory of relativity is built based on the
assumption that the choice of the coordinates is not essential for the
statement of the law of bodies movement (This looks rather  like our
freedom to choose the units in equations (3-5)). Quantum
electrodynamics can be built on the assumption that the wave function
phase can not be physically seen and its selection is arbitrary. Theories
of weak and strong interactions also have fundamental symmetries of
selection of physically equivalent objects. Physicists refer to such
symmetries as gauge symmetries. The theory of the financial market
described in this paper can be also called a gauge one - the Gauge Theory
of Arbitrage (GTA) since the role of the force field  in this theory is
played by the excess return on arbitrage operation. From a mathematical
point of view, this theory looks very much like electrodynamics with only
one difference between them: instead of a local group of quantum phase
rotation a local group of dilatations of financial assets units is used.

\subsection{GTA Geometry (can be skipped when read for the first time)}

Using a more formal approach one can assume that the inner reason for the
similarity between all contemporary physical theories and the GTA lies in
the fact that all of them are descriptions of dynamics in geometrically
complex spaces called by mathematicians fibre bundles. Such spaces
consist of a base $B$ (for instance, Minkowski space or only two straight
lines D and P) and fibres $F$ adhered to each point $B$. Such fibres
collected in one point make a bundle, which accounts for a generic name
of such spaces. Imagine that we are watching a particle moving in such a
fibre bundle. The particle can move inside the fibre as well as between
fibres so that its position can be described by two characters ($x$ and
$y$):  $x$ refers to the particle coordinate on the base $B$, while $y$
denotes the particle coordinate in the fibre $F(x)$ corresponding to
point $x$.  Putting aside the issue of coordinates on the base we
can ask a question:  "What if coordinates in different fibres are not
adjusted to each other?". In this case changing of coordinates does not
say anything about the real change of particle position which is
characterized by change of "real" coordinates -- exactly like the rate of
return does not say anything until inflation is accounted and the real
rate of return is calculated by the Fisher's formula. Like in the example
with inflation, we must subtract from the total change of the coordinate
its superficial change which is associated with zero real change and
which is determined by coordinates disagreement in different fibres. It
is this superficial change that determines the rule for coordinates
comparison in coordinate systems of different fibres or, as
mathematicians put it, parallel transport.

Going back to the main topic of our discussion we can see that the
abstract discourses described above look as if they were specially
designed for the financial market.  Assume we have two currencies (two
points on the base) and want to compare four dollars and three pounds
(numbers 3 and 4 in coordinate systems of fibres corresponding to points
"dollar" and "pound" on the base). At first glance fours notes seem more
attractive than three. However, anyone will prefer to have three pounds
because at an exchange rate 1.67 dollars per one pound one can get five
dollars for these three pounds, which is much better than to have only
four. We can see that when assets transformed from four dollars into
three pounds the real change was equal to +1 dollar instead of initial
 -1. In our case the real value of four dollars was 2.40 pounds, and
the fictitious change was equal to 1.60 pounds. When compared 2.40 pounds
differ from three pounds by 60 pence, which is equivalent to 1 dollar
exactly. For a mathematician all these would mean that 2.40 pounds are
equal to 4 dollars under parallel transport from one point on the base to
another, and a covariant (real) difference of three pounds and four
dollars is equal to 60 pence.

Net Present Value gives us another financial example of parallel
transport. Assume that we can choose between 100 pounds now or 103 pounds
a year later. At first glance one hundred and three (the same currency!)
notes seem more attractive than one hundred. However, a reader is likely
to choose one hundred because at an interest rate 5\%
one hundred pounds
now will become 105 pounds in a year, which is definitely better than
103. Instead of counting all pounds a year later we can count them now.
Then, we will be able to compare 100 pounds with the discounted by
present value of 103 pounds, that is the Net present Value (NPV) equal to
98.10=103/(1+0.05). Again one can see that as the assets move in time
from pounds at present to pounds a year later the real change is –2
pounds instead of initial +3, the parallel translation of 100 pounds
amounting to 105 pounds and the covariant (real) difference being –2.

After we have defined parallel transport, we can address an issue of the
difference of results of parallel transport carried out by different
routes on the base which have common beginning and end. However, from
technical point of view it is more convenient to deal not with a pair of
routs but with one cyclic path that covers one route moving forward and
the second one moving backwards. The difference between the initial value
of the transported amount and its value after the parallel transport
along the closed path determines the curvature tensor of the fibre
bundles, and, as it follows from formulas (3) and (4), directly results
in excess rate of return on arbitrage operation and preconditions the
assets movement. That is why comparison of the excess rate of return with
the electromagnetic field intensity is not accidental – both make
particles move and both are equal to elements of the curvature tensor! In
physics different results of parallel transport always mean the affect of
forces on the transported particle, and the value of these forces is
equal to elements of the curvature tensor. Introduction of the value R in
formula (5) is not accidental as well – it is directly associated with
the square of  the curvature tensor which is equal to the field energy
both  in our example as well as in electromagnetic field.

To sum up, we can say that when financiers buy and sell securities,
exchange currency or calculate NPV, they make parallel transports in
fibre bundles. What is more, they have been making that for more than 100
years without knowing about it. It is like Moliere's famous hero who was
very surprised when he learned that he spoke prose. The absence of
arbitrage and "free lunches" looks very much like the condition of energy
minimization, which is ubiquitous in classical physics.

\subsection{Uncertainty and Quantization}

The picture we have built is based on the assumption that all information
capable to affect prices is known in advance, which means there is no
uncertainty. The real financial world, as you understand, is far from
this picture. Any financial operation is associated with a risk immanent
to the financial world, that is why we have to generalize the theory in
case of uncertain or random prices.

Describing random characters we are not able to predict their exact value
such as the exchange rate, for example, in a month time. What we can do
is only predict the probability with which we can expect a certain value
of the exchange rate. It is difficult to define the very concept of
probability in this case. According to the definition, probability is
equal to the number of times a certain result is expected to be achieved
under identical experimental conditions. This means that applying
probabilistic description we expect that the experiment will be repeated
in the exactly the same way many times. It is evident that this condition
can not be met in the example with the exchange rate: there is only one
April and only one May in 1998. Trying to repeat the experiment in
May-June we will encounter different conditions on the market, and
consequently, we will have to change the conditions of the experiment.
Moreover, the result of the experiment in April-May may affect the
outcome of the experiment in May-June. We will try to overcome these
difficulties using conditional probabilities and short periods of time
when it can be assumed that all outer factors remain unchanged.

Like random values, random paths are also described by probabilities, in
this case – the probability of covering the whole path. Such a
probability is referred  to as the path weight. The weight fully defines
a statistical model and the choice of  the weight is a key step in
building a theory. Here again we are facing the problem of choice. To
make our choice we will use the following considerations:  The weight
must keep the symmetry of the theory built before stochasticity could
enter it, that is the weight must be based on gauge-invariant quantities
like (3), (4) and (5).  In the situation when the world becomes less and
less accidental, the weight must identify paths with minimum mispricing
opportunities, that is the weight must identify the paths on which the
quantity (5) has a minimum value.  In the approximation where the speed
of money flows is infinite and they do not effectively participate in the
description, the theory, according to equation (1), must reproduce the
results of financial mathematics.  Traders represent a homogeneous
ensemble. Such an assumption distinguishes this model from the previously
proposed ones that represented the market as a combination of "smart
money" and "noise traders". "Smart money" know the real price and remove
mispricing, while "noise traders" behave unreasonably and represent a
crowd. Remember that our target is to describe the short-term dynamics
when all traders are professional participants of the market, and it
would be odd to divide them into "smart" and "noise". It is more
realistic to assume that all traders behave in a different way in terms
of individuality but in the same way in terms of statistics, and their
understanding of the real price is formed according to a stochastic
distribution. The latter must be gauge symmetric. Of course, a question
how to model traders' behavior still remains. In other words, it is
unclear whether an assumption about return maximization is sufficient or
one should introduce additional factors such as risk aversion, herd
behavior, etc. This question will be addressed in the next paragraph.

It can be demonstrated that a simple theory built on the basis of these
considerations will be as similar to quantum electrodynamics (in
imaginary time) as a theory without stochasticity is similar to the
classical electrodynamics.

\section{Answering the Three Questions "Why?"}

After having described the general principles of construction of the
theory we will consider its applications and see how this theory answers
the three questions "Why?" of financial physics put in the first
paragraph.

\subsection{Statistical Characteristics of Real Price Charts}

If one suggests that price changes are associated with new fundamental
information, the chart of a return must look like the path of the
particle movement affected by a random force. After some time, the return
value may be defined with some probability, and the distribution function
of this probability will be a Gaussian function, which means return will
have normal distribution. However, it was noticed long ago that real
profitableness distribution functions do not look like Gaussian
distribution: at the same average values and variation they have "fat"
tails and a higher and narrower peak. In addition, the peak height comes
down eventually not according to a conventional distribution law
$t^{-0.5}$ but according to the law $t^{-a}$ where exponent $a$ depends
on the concrete security and is often approximately equal to 0.7. The
latter is called by physicists an anomal scaling behavior and is
associated with a nontrivial internal dynamics. The description of this
dynamics will be an answer to the first question "Why?".

The scaling behavior can be explained by the Fractional Market Hypothesis
(FMH) (see Peters, 1995 for discussion and motivation) according to
which, in the market there are investors with different investment time
frames but the same general laws regulating investors' behavior which do
not depend on the time frame. Should the FMH be accepted, it is
sufficient to describe the behavior of investors with short time frames
to obtain the description of the system in general. It is here that the
gauge theory comes on to the stage.

Fig. 1 and 2 show the results of a statistical study (Mantegna and
Stanley, 1995) of the price distribution function of a portfolio of
stocks belonging to 500 largest companies (squares) traded on the New
York Stock Exchange (market index S\&P500) and a model distribution
function (a solid line) achieved from a simple model of the Gauge Theory
of Arbitrage (Ilinski and Stepanenko, 1998). Fig. 1 shows the law of the
distribution function height descent with time (the anomal scaling law),
while Fig. 2 demonstrates the function distribution  at one minute. While
the scaling behavior owes its appearance to the FMH, the distribution
function type was defined due to statistical laws of gauge dynamics. It
is a nearly ideal matching of the theory and the "experiment", which is
similar to the accuracy of quantum electrodynamics. Fig. 2 also shows
normal distribution (long dashes) and the popular Levi distribution
(short dashes) having the right scaling behavior. Both of them are not as
accurate in description as the model distribution.

\subsection{Technical Analysis and the Efficient Market}

A conventional simplified understanding of an efficient market is based
on the assumption that all relevant information is already reflected in
the price, and the price may change only due to some news. Since the news
can not be forecast by virtue of its definition, the price changing is a
random process which is not affected by the already known information,
say price history. Thus, technical analysis using price history for
forecasting future prices falls out of the law, and a trader instead of
following a system approach in trading is resorted to toss a coin. It is
unlikely that she or he will accept this suggestion.

The cause of this seeming contradiction lies in the unclear statement of
the efficient market hypothesis. An accurate statement calls for
rationality from investors and availability of a real pricing model and
assumes that the real price may deviate of from the model only
occasionally. The model price itself can depend on time and be defined by
the available information, for instance, previous prices. Let us consider
the following example. Assume that due to new information the return of
certain securities has been increased compared to other securities with a
similar risk. More profitable securities will be purchased until the
return is leveled. The process of leveling takes time, and it should be
accounted for in the construction of an ideal model. More details about
equilibration can be taken from price history. If this washing out of the
mispricing takes finite amount of time, the price history analysis and
the market forces behind it may become a key factor in building a model
of the dynamics of future prices or, strictly speaking, of their average
values. The latter brings us back to technical analysis representing a
set of empirical (a physicist will say "phenomenological") rules for
prediction of the model prices and making respective investment
decisions. Comparison of the characteristic time of return fluctuation
and the time of relaxation determines the applicability of technical
analysis. The assumption about immediate relaxation leads to the initial
simplified definition of the efficient market.

All said above can be applied to the Gauge Theory of Arbitrage. In the
previous paragraph it was demonstrated how a stochastic description
appears in the gauge theory. The aim of this paragraph is to show how
technical analysis emerges in the same framework. To a certain extent,
technical analysis plays the role of classical mechanics on which the
quantum theory is founded. Within relatively short time frames any
quantum dynamics comes down to the classical one -- if a quantum particle
is released and its behavior is observed within a short time frame,
although it will be impossible to define its exact location, the degree
of uncertainty of its location will be negligible and its mean value will
be the same as the result of a solution of the classical  equations of
motions (that is why such an approximate description is referred to as
quasi-classical). Uncertainty grows with time, and a quasi-classical
solution degrades and deforms. Stochastization takes place.

It can be demonstrated that in a quasi-classical approximation (for short
time frames) a simple gauge model used in the previous paragraph to find
the distribution function come down to a system of equations the solution
to which corresponds to two known technical analysis indices – the
Negative Volume and the Positive Volume Indices. These indices describe
the behavior of rational and irrational investors, respectively. The fact
that both indices have appeared in one model is not a mere coincidence –
the statistical behavior of traders was defined as an irrational
distribution of solutions around a rational average one, and that is why
in quasi-classical equations both a rational and a noise  components are
found. It seems logical to refer to a market of such agents as a
quasi-efficient market.

To sum up, technical analysis in the gauge theory corresponds to
quasi-classical dynamics and describes dynamics within short time frames
which are followed by stochastization resulting in a realistic
statistical description. The question why technical analysis can be
applied to different investment time frames can be answered by the
Fractional Market Hypothesis.

\subsection{Portfolio Theory and Derivative Pricing}

A term "investment portfolio" is a key term in a contemporary financial
theory. In assessing derivatives a portfolio replicating a derivative
plays an important role and determines the cost of the derivative. In the
Asset Pricing Theory an optimal portfolio defines the average asset
growth as a function of its correlation with the key market parameters.
In both cases prices are considered to be a random process, and the task
comes down to elimination and averaging stochasticity and assessing
securities under the condition of absence of arbitrage opportunities. The
next step to be taken should be setting up a portfolio theory which
would, on the one hand, accept temporary arbitrage opportunities (they
always exist in the market) and, on the other hand, provide for
elimination of such opportunities and allow for accounting a no-random
component of pricing movement and the technical analysis predictions. All
these tasks can be addressed within the frames of the Gauge Theory of
Arbitrage.

As it was demonstrated in previous paragraphs, the gauge approach
provides for description of the market response to appearance of
arbitrage opportunities and relaxation caused by it. It also allows to
model the realistic price behavior and apply technical analysis tools
having selected an adequate classical theory. One of the applications of
this approach is finding adjustments to the Black-Scholes equation when
virtual arbitrage opportunities and complicated pricing processes are
present. To this end, two sorts of traders should be introduced into the
theory – speculators and arbitrageurs. The former are involved in
modeling a realistic pricing movement, while the latter account for
diminishing of arbitrage opportunities. Such classification does not seem
artificial, and can be virtually seen in the market. In the frames of an
infinitely fast response of investors, as one might have expected, the
task can be limited by the solution of the Black-Scholes equation.
However, within limited time of relaxation the gauge theory generates
adjustments to the equation or, strictly speaking, makes it an
integro-differential due to the memory effects. The same approach can be
also applied for deriving adjustments to the equations of the Arbitrage
Pricing Theory and, eventually, to the Capital Asset Pricing Model
bearing in mind that in the latter case arbitrage opportunities are
likely to appear not between a derivative and a replicating portfolio but
between two different portfolios.

Finally, one more comment should be made. Besides the effects described
above, a gauge approach has one more advantage – it allows the inclusion
of such market imperfections as bid-ask spread and transaction costs. To
introduce them into the model one has just to insert additional factors
into the matrix of model transitional probabilities.

\section{Conclusion}

The aim of this paper was to acquaint readers with a number of new
concepts and tasks arising when analyzing the influence of money flows on
the random pricing process. To a certain extent all scenarios described
above were known earlier, and we have just made an attempt to give an
idea of a quantitative description of the processes on which these
scenarios are based. Technical issues were not addressed in order to make
the principles of the theory clearer.

Although from the formal point of view the theory is similar to quantum
electrodynamics, it is unlikely to be as fundamental as its prototype. It
is rather a new language convenient for formulating the problems and
searching for their solution. This language is technically advanced, and
we hope that we have managed to persuade the readers, at least partly,
that it is adequate and useful. Using the terms of this language one can
introduce new elements relevant for individual problems (such as behavior
peculiarities of investors or market restrictions), which will allow to
extend the range of problems that can be modeled.

In conclusion, we can give an example of an algorithm for building a
pricing model of a concrete instrument within the framework of the gauge
approach. The first step is to select a classical gauge-invariant theory
for money flows which correlates with a certain tool of technical
analysis applicable to this particular sort of securities. Then,
following a formal prescription one should introduce a stochastic factor,
which means to construct a quantum theory. The next step is to verify the
validity of the stochastic description of the model on historical data.
By this time all the parameters of the theory will have been determined,
and the pricing model can be considered completed. After that, the tested
model can be used by investors with different purposes, for instance, for
arbitrage, speculation, hedging and risk management.

\section*{References}
\begin{enumerate}
\item
K. N. Ilinski: Physics of Finance, to appear in {\it Econophysics:
an emerging science}, Dordrecht: Kluwer (1998);
available at http://xxx.lanl.gov/abs/hep-th/9710148;
\item E. E. Peters, {\it
Fractional Analysis}, John Wiley \& Sons, Inc., 1995;
\item R. N. Mantegna
and H. E. Stanley:  Scaling behavior in the dynamics of an economical
index, {\it Nature}, {\bf 376}, 46-49, 1995;
\item K. N. Ilinski and A. S.
Stepanenko: Electrodynamical model of quasi-efficient market, to appear
in {\it J.Complex Systems} (November 1998);
http://xxx.lanl.gov/abs/cond-mat/9806138.
\end{enumerate}

\newpage

\begin{center} {Figure caption} \end{center}

\vspace{2cm}

FIG.1 Theoretical (solid line) and experimental (squares) probability of
return to the origin (to get zero return) $P(0)$ as a function of time.
The slope of the best-fit straight line is -0.712$\pm$0.025.
The theoretical curve converges to the Brownian value $0.5$ as time tends to
one month.

\vspace{2cm}

FIG.2 Comparison of the $\Delta =1$ min theoretical (solid line) and
observed (squares) probability distribution of the return
$P(r)$.  The dashed line (long dashes) shows the gaussian distribution with the
standard deviation $\sigma$ equal to the experimental value 0.0508.
Values of the return are normalized to $\sigma$.  The dashed line (short dashes) is the
best fitted symmetrical Levy stable distribution.

\begin{figure}
\centerline{\epsfxsize=16cm \epsfbox{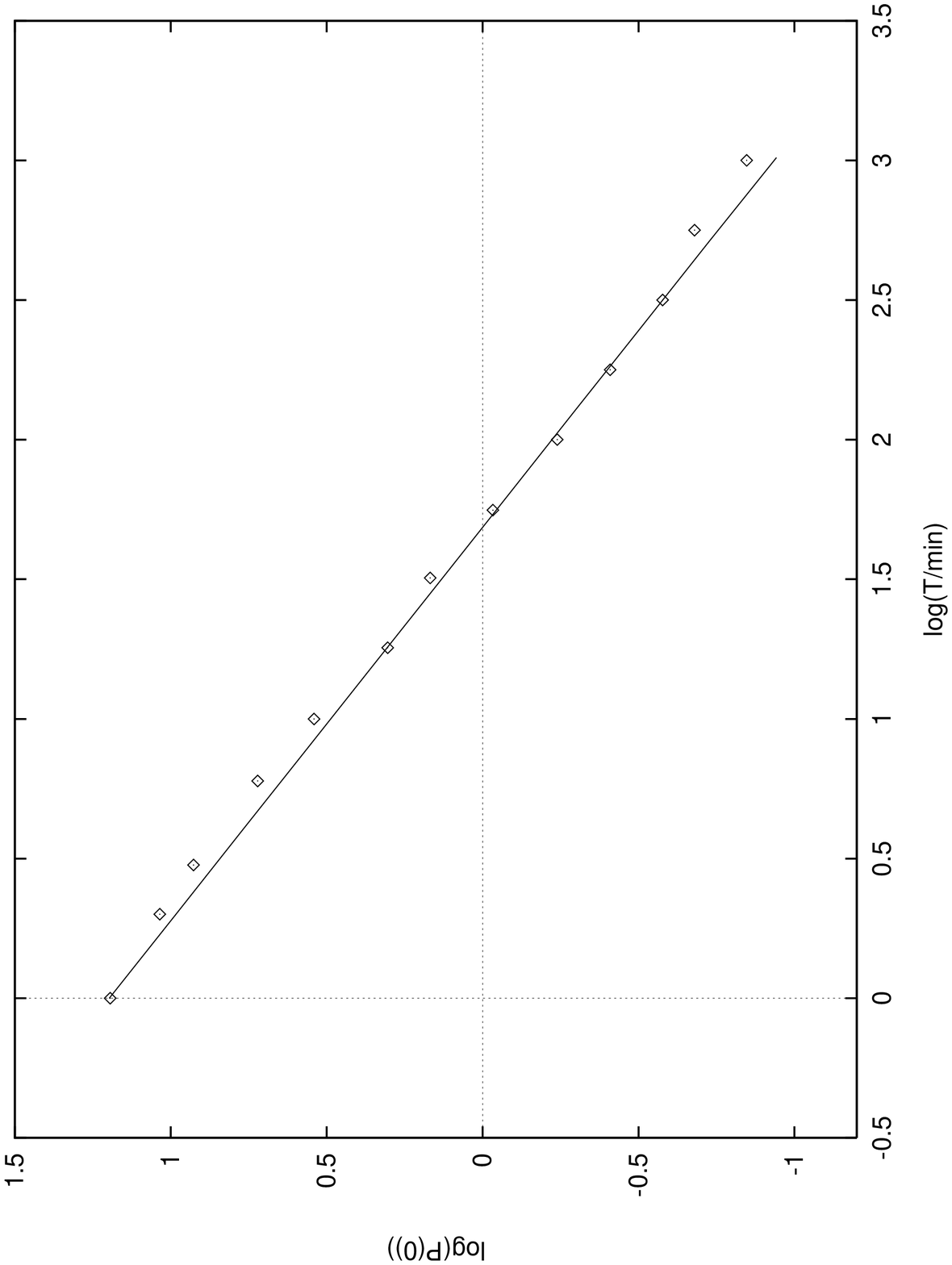}}
\end{figure}

\begin{figure}
\centerline{\epsfxsize=16cm \epsfbox{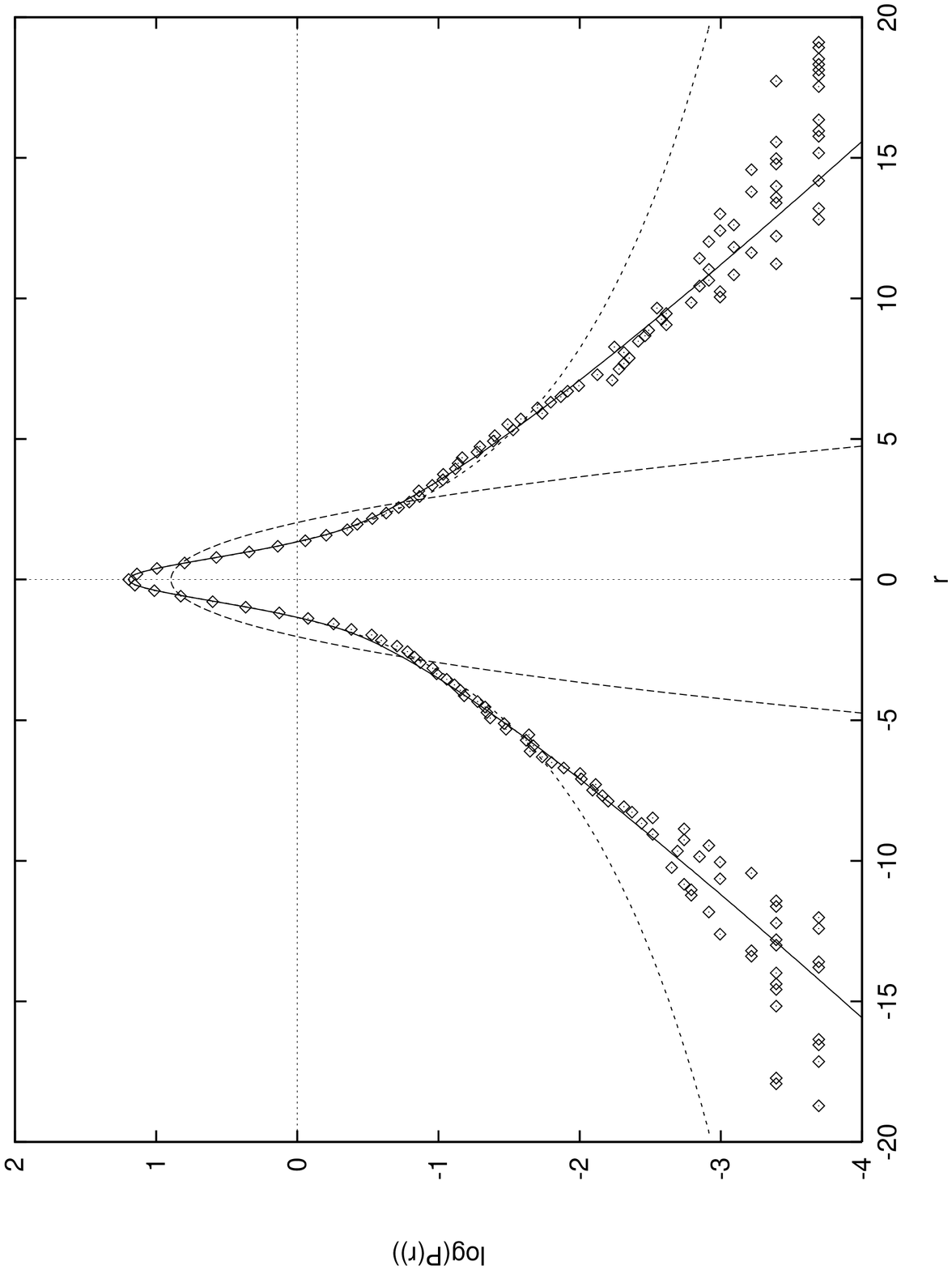}}
\end{figure}

\end{document}